\documentclass{aastex}
\usepackage{spr-astr-addons}

\RequirePackage{color}

\begin{document}

\title{Signatures of recent star formation in ring S0 galaxies}
\slugcomment{Not to appear in Nonlearned J., 45.}
%% Running heads
\shorttitle{Signature of recent star formation in ring S0 galaxies}
\shortauthors{Marino et al.}

\author{A. Marino, L. Bianchi}    
\affil{Dept. of Physics \& Astronomy\\
Johns Hopkins University \\
3400 N. Charles St.\\
 Baltimore, MD 21218}
 \and 
\author{R. Rampazzo}
\affil{INAF Osservatorio Astronomico di Padova \\
Vicolo dell'Osservatorio~5 \\
I-35122  Padova, Italy}
\and 
\author{D. Thilker}
\affil{Dept. of Physics \& Astronomy\\
Johns Hopkins University \\
3400 N. Charles St.\\
 Baltimore, MD 21218}
\and 
\author{F. Annibali, A. Bressan, L.M. Buson}
\affil{INAF Osservatorio Astronomico di Padova \\
Vicolo dell'Osservatorio~5 \\
I-35122  Padova, Italy}

%\author{A. Marino, L. Bianchi, D. Thilker}
%\affil{Dept. of Physics and Astronomy \\
%        Johns Hopkins University\\
%         3400 North Charles Street\\
%         Baltimore, MD 21218  USA}
%\email{\emaila amarino@pha.jhu.edu}
%\and \author{R. Rampazzo, F. Annibali, A. Bressan, L.M. Buson}
%\affil{INAF-Osservatorio Astronomico di Padova \\
%        Vicolo dell'Osservatorio 5
%        35122 Padova, Italy }

\begin{abstract}
We present a study of the stellar populations of ring and/or arm-like 
structures in a sample of  S0 galaxies using {\it GALEX} far- and near-ultraviolet 
imaging and SDSS optical data.  
Such structures are prominent in the UV and reveal recent star formation. 
We quantitatively characterize these rejuvenation events, estimating the average age 
and stellar mass of the ring structures, as well as of the entire galaxy.
The mass fraction of the UV$-$bright rings is a few percent of the total
galaxy mass, although the UV ring luminosity reaches  70\%
of the galaxy luminosity. The integrated colors of these S0s locates them  in the red
sequence (NGC 2962)  and in the so-called green valley.  We suggest that the 
star formation episodes may be induced by different triggering mechanisms,
such as the inner secular evolution driven by bars, and interaction episodes.

\end{abstract}


\section{Introduction}

S0 galaxies have been introduced by \cite{Hubble36} 
in his ``tuning fork" galaxy classification as a more or less hypothetical 
transition class between ellipticals and spirals. The bulge and the disk 
are the defining structures of  S0s. Later classification schemes by 
\citet[][RC3 hereafter]{RC3} and \citet[][RSA hereafter]{RSA} 
 take into account the presence of the several sub-structures detected
within the S0s class: bars, inner and outer rings, as well as lenses
(from here the widely used term of lenticulars as synonym of S0s) 
and ovals are often found.   

% -----------------------------------------Figure 1
\begin{figure}%[tb]
\includegraphics[width=\columnwidth]{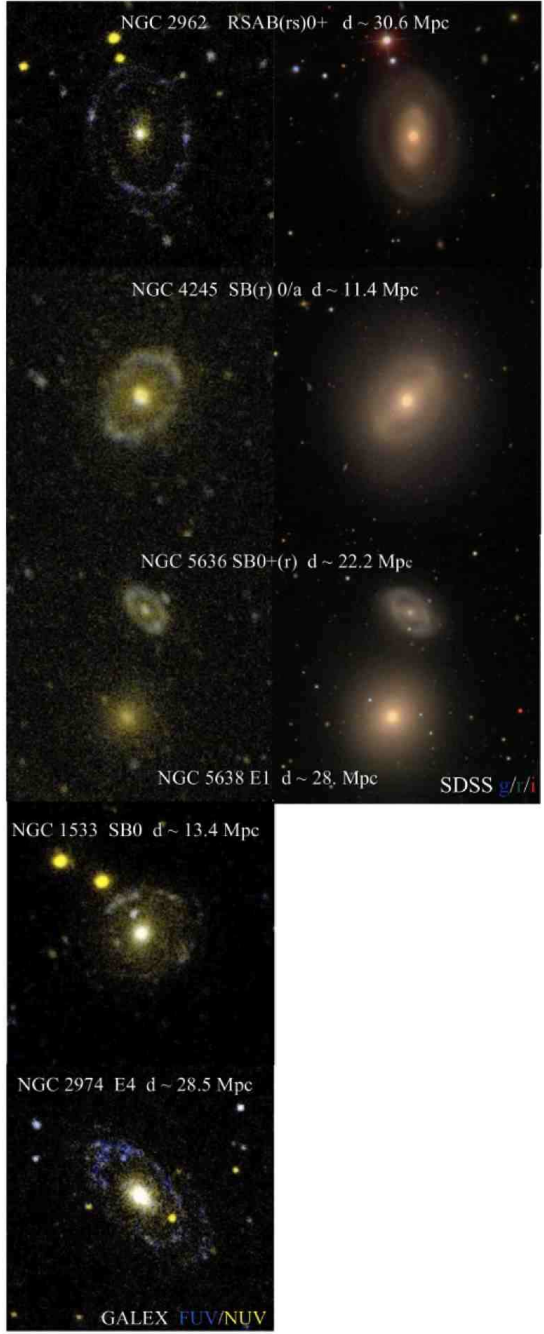}
\vspace{-0.2cm}
\caption{Composite {\it GALEX} images (FUV=blue and NUV=yellow; left panels) and 
SDSS ($g$=blue, $r$=green, $i$=red; right panels) color composite images 
of the S0 galaxies NGC 1533, NGC 2962, NGC 2974, 
NGC 4245 and NGC 5636  and of the elliptical NGC 5638. 
The outer ring structures appear in the UV less smooth than in optical 
(in spite of the SDSS higher resolution) and  bluer than the nucleus, 
suggesting the presence of young stellar populations}
%% no full stop at the end of caption
\label{fig1}
\end{figure}
%----------------------------------------- end Figure 1

S0s are,  as a rule, admitted in the vast class of early-type galaxies 
(ETGs hereafter),  together with Ellipticals, with which they share typically 
passively evolving stellar populations. From an evolutionary point of view
it is  widely believed that S0s were initially spirals which lost their interstellar
medium (ISM hereafter) during collisions \citep{Spitzer51} 
by ``harassment'' \citep{Moore96} or by
``ram-pressure''. Simulations show that such events could occur both in 
clusters \citep{Gunn72} and in groups  \citep[see e.g.][]{Bekki09}. 

From the kinematical point of view, genuine S0s are quite distinct 
from giant ellipticals since they are fast rotating, like late-type galaxies.
At the same time, if S0s were originally spirals, they should have formed 
their mass over a significant fraction of the Hubble time, following a star formation
history more similar to that of late-type  galaxies rather than that 
of giant elliptical galaxies.
In this context,  our multi-wavelength study of nearby ETGs 
\citep{Marino10a} shows that S0s, characterized through their 
luminosity profiles and by the low values of the $n$
index of the Sersic law \citep{Sersic68}, have the  lowest values  
of  the $\alpha-$enhancements ($[\alpha/Fe]$) in the sample. The low
$[\alpha/Fe]$ values  suggest a ``more prolonged'' star formation 
in the S0s galaxies with respect to ellipticals in the sample 
\citep[][see also Rampazzo et al. these proceedings]{Annibali07,Annibali10}.
 
Several mechanisms, both internal and external to 
the galaxy, may be envisaged to produce the signature
of a ``prolonged'' star formation.  
The removal of the ISM  from a possible spiral precursor
may quench the ongoing star formation transforming 
the debris in an S0 with signatures of prolonged 
star formation, as well as an external ``wet'' accretion 
and/or the inner secular evolution, e.g. driven by a tumbling bar
\citep[see e.g.][]{vandenbosch98}.     

The {\it Galaxy Evolution Explorer} satellite ({\it GALEX} hereafter)
has widely contributed to explore the above mechanisms. 
``Wet'' accretions, and the rejuvenation of the stellar population,
has been evidenced in ETGs \citep[see e.g.][and Rampazzo et al. in these
proceedings]{Marino09}.
Recently, \citet{Thilker10} found in NGC~404,
a nearby well known S0, an external ring-like structure with signature
of recent star formation. 70\% of the FUV comes from an HI ring  forming
stars at a rate of 2.5$\times$10$^{-3} M_{\odot}$ yr$^{-1}$.  The structure
has been likely produced by a ``wet accretion/merger''.
Along this line, using {\it GALEX} we analyze five S0s, showing outer rings and/or 
arm-like structures, from the sample of  \cite{Marino10a, Marino10c}
aiming at understanding the nature of the ring and to map possible
star formation episodes.

  % -----------------------------------------Figure 2
\begin{figure*}[!t]
\centering \includegraphics[width=12.2cm]{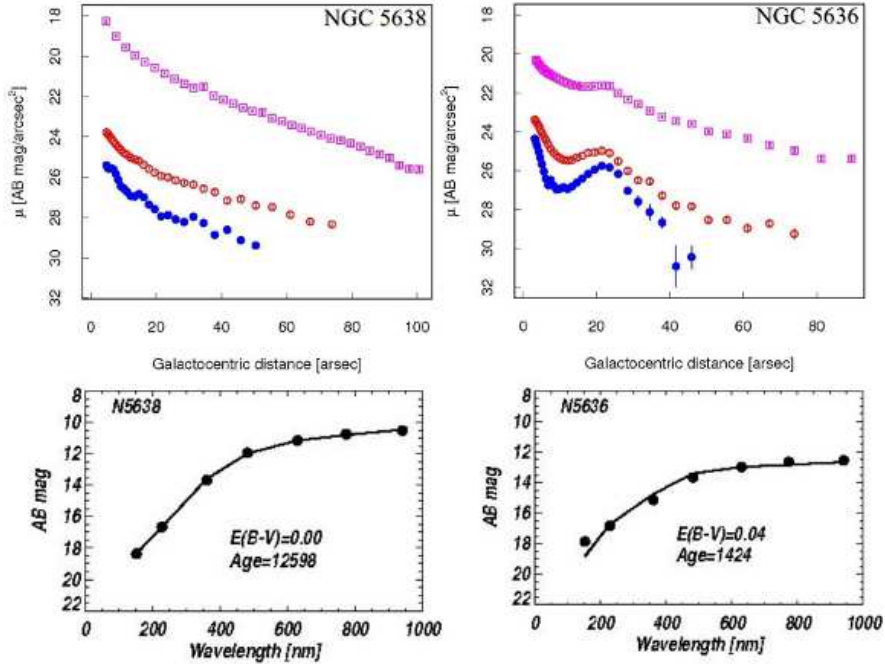}
 \caption{In the top right panel the optical and UV 
surface brightness profiles of NGC 5636 are plotted. 
Blue filled circles represent FUV, empty red 
circles   NUV, and magenta empty squares SDSS$ r$.
For comparison, the smooth FUV, NUV and $r$ surface brightness profiles 
of the elliptical NGC 5638 are shown in the top left panel. 
In the bottom panels the full dots represent the FUV, NUV,  $u$,
$g$, $r$, $i$, $z$ spectral energy distribution of these galaxies. Best fit models,
assuming only foreground extinction, are indicated with continuous lines (see text).
Ages, derived from SED fitting, are reported in Myrs} 
%% no full stop at the end of caption
\label{fig2}
\end{figure*}
%----------------------------------------- end Figure 2

\section{The sample}

The sample includes five S0 galaxies: NGC 1533,  NGC~2962,  NGC~2974,  
NGC~4245 and NGC~5636 (interacting with NGC~5638, an elliptical galaxy). 
Notice that NGC~2974 is erroneously classified as elliptical  
both in RC3 and RSA, although an exponential
disk is evident from the optical and UV luminosity profiles 
\citep[see also][]{Jeong07, Jeong09,
Marino10a}. In addition, in NGC 2974 \citet{Krajnovic05}
found, with {\tt SAURON} observations \citep{deZeeuw02}, 
the existence of non-axisymmetric 
perturbations consistent with the presence of inner bars.
Summarizing, all galaxies in the present sample are barred S0s. 

In some of these galaxies the literature reports signatures of
both possible recent accretion events and of ongoing interaction. 
\citet{Tal09} report the presence of a shell system surrounding NGC 2974,
a signature of a minor merging event \citep{Dupraz86,Ebrova09}.   
\cite{Werk10} detected a large HI structure (see her Figure~7) extending 
for several kiloparsec outside NGC 1553 towards the interacting companion, IC 2038.
In the HI cloud, well outside the NGC 1553 ring,  OB associations are also found.
\citet{deGraaff07} describes this galaxy as a late stage of 
transition from a barred spiral to a barred S0 galaxy.

All S0s belong to low density environments since they are members 
of loose groups of galaxies. NGC 1533 is member of the Dorado group.   
NGC 2962, NGC2974,  NGC 5636 and its elliptical companion NGC 5638,
belong to  LGG 178, LGG 179, LGG 386 groups, respectively \citep{Garcia93}.
NGC 4245 is located in the spiral-dominated group USGC-U478 \citep{Ramella02}.

\section{Observations}

UV imaging was obtained with  {\it GALEX} \citep{Martin05,Morrissey07} in 
far-Ultraviolet (FUV, 1344 -- 1786 \AA) and near-Ultraviolet 
(NUV, 1771 -- 2831 \AA) bands. 
The journal of the {\it GALEX}  observations is given in Table~1.
In Figure~1 we show the composite UV and optical images of the galaxies. 

Ancillary optical  data for NGC 2962, NGC 4245, NGC 5636 
were also retrieved from the SDSS archive  \citep{Ade08} in
the $u$ [2980-4130 \AA], g [3630-5830 \AA], $r$ [5380-7230 \AA], $i$
[6430-8630 \AA] and  $z$ [7730-11230 \AA] bands. 

 %----------------------------------------------Table 1 ----------------------
\begin{table}[!h]
\scriptsize{
\caption{Journal of the {\it GALEX} observations}
\begin{tabular}{lllllllll}
\hline\hline
\multicolumn{1}{l}{Ident.}&
\multicolumn{1}{l}{FUV }&
\multicolumn{1}{l}{NUV } &
\multicolumn{1}{l}{Observing} \\
%\multicolumn{1}{l}{PI} \\
\multicolumn{1}{c}{} &
\multicolumn{1}{l}{Expos.} &
\multicolumn{1}{l}{Expos.} &
\multicolumn{1}{l}{program} \\
\multicolumn{1}{c}{} &
\multicolumn{1}{l}{[sec]} &
\multicolumn{1}{l}{[sec]} &
\multicolumn{1}{l}{} \\
\hline
 NGC 1533 &1520 & 3152 &GI3 087004      \\ 
 NGC 2962& 2297 & 2297 & MISWZN09\_24136\_0336 \\  
 NGC 2974& 2657 & 2657 & GI1\_109006    \\ 
 NGC 4245 & 1680 & 1680 & GI3\_041006\_NGC4245\_0001   \\
 NGC 5636 & 1704 & 1704 & MISDR1\_33739\_0535 \\
 \hline												    
\end{tabular}}											    
\label{table1}											    
\end{table}											    
%------------------------------end Table 1------------------------------------- 		    

\section{The UV-bright outer rings}

NGC~1533 UV image shows  an incomplete, clumpy ring-like structure,  
while NGC~2962, NGC~4245 and NGC~5636 which have
well defined and complete rings, although not uniform, in the FUV and NUV
bands.  Spiral arm-like structures are visible in both NGC~2962 and in NGC~2974.
Both the ring and the arm-like structure appear bluer than the  nucleus 
and the galaxy body. The bar present in the optical images of the galaxies
are barely visible in the FUV and NUV  images. 
In all but NGC 2962 and NGC 5636 rings are within the $D_{25}$ isophote. 

In Figure~\ref{fig2} (top panel) we show the FUV,  NUV and  $r$ bands 
surface brightness profiles  of NGC 5636 and, for comparison, 
the smooth profiles of the elliptical companion NGC 5638, 
obtained with the {\tt ELLPISE} 
routine in the {\tt STSDAS} package of {\tt IRAF}.
Rings produce a hump in the surface brightness profiles, 
more prominent in FUV and NUV than in 
optical. Outer rings  are  bluer than  the body of their 
respective galaxies indicating  the presence of young 
stellar populations.

We compute FUV, NUV and optical magnitudes 
of the outer ring structures using surface brightness profiles
after subtracting the galaxy contribution as  
described by \citet{Marino10b}.
About 25\%, 71\%, 30\%, 33\% and 60\% of the total FUV luminosity of 
NGC 1533, NGC 2962, NGC 2974, NGC 4245, NGC 5636 comes from 
the ring structure.  

We also compute the total galaxy magnitude  in the AB system integrating
the sky-subtracted UV and optical galaxy light within concentric elliptical apertures 
enclosing the galaxy.  This aperture photometry provides the FUV to 
near-Infrared spectral energy distribution (SED hereafter) of our galaxies. 
In Figure~\ref{fig2} we plot, as an example, the SEDs of NGC 5636 and 
of the companion elliptical galaxy NGC 5638 (see next section). 

\section{Characterizing the star formation history}

We aim at characterizing the strength and the epoch of the rejuvenation episodes in
S0s by estimating the ages and stellar masses of the outer rings and of  the entire 
galaxy. 

To this purpose, we compare the observed SEDs with population synthesis  
models computed with the {\tt GRASIL} code \citep{Silva98} which takes into 
account the effects of the internal extinction.  The star formation history (SFH) and
 the Initial Mass Function (IMF), the metallicity and the residual gas 
fraction are given as input parameters of the code. 
We %compared the observed SEDs with grid of theoretical SEDs obtained  
adopted two different  SFHs: one %which we consider
typical of elliptical galaxies, characterized by a short (1 Gyr) and 
intense period of star formation followed by pure passive stellar evolution,  
and the other one, typical of spirals with a more prolonged SF.
We used  a Salpeter IMF with a mass range between 0.15 and 120 M$_{\odot}$
for ellipticals and between 0.1$-$100 M$_{\odot}$ for spirals 
following \citet{Silva98}.  A gas time infall 
$t_{inf}$=0.1 and 4 Gyr and a star formation efficiency $\nu$=2 and 0.6 
have been used for elliptical and spirals respectively. 
The code computes the integrated spectrum taking into account the stellar
and gas budget at any age. It includes the effect of the age-dependent
extinction with young stars being more affected by dust.

% -----------------------------------------Figure 3
\begin{figure}[!h]
 \includegraphics[width=\columnwidth]{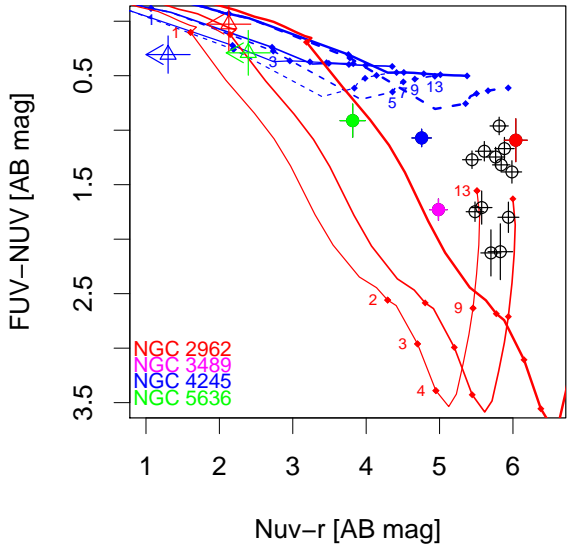} 
   \caption{Top: (NUV$-r$) vs (FUV$-$NUV) measurements of outer ring 
structures (triangles), and  total (colored filled circles) of NGC 2962, 
NGC 5636, NGC 4245. We also plot NGC 3489, an S0 in the sample of 
\citet{Marino10a}, showing blue arm-like structures. 
Black open circles plot ETGs in \citet{Marino10a} without ring structures. 
Blue lines represent {\tt GRASIL} model colors for  disk-type SFH with different 
 extinction (E(B$-$V) = 0, 0.1, 0.3 from thin to thick lines)  and inclination 
 (solid line i=0$^\circ$, dashed line i=90$^\circ$).  Red lines are {\tt GRASIL} 
 model colors for E-type SFH (passive evolution).   
   The UV-rising part in E-type models at old ages  is due 
 to post-AGB  stars. Numbers on the model lines are ages in Gyrs 
}%% no full stop at the end of caption
\label{fig3}
\end{figure}
%----------------------------------------- end Figure 3

In Figure~\ref{fig2} (bottom panels)  we plot SEDs of  NGC 5636 and
NGC 5638,  obtained  integrating  the light of the entire galaxy, and 
the best-fit model obtained adopting an Elliptical SFH. 

The elliptical NGC 5638 is  well fitted with an  old
age (12 Gyr) of the stellar population. 
The case of the ring S0 companion NGC 5636 is quite different: the observed
SED shows  a FUV excess with respect to the best-fit model assuming  
foreground extinction of 0.04 indicating that a simply passive SFH is inappropriate.  
%Leaving to variate  E(B-V) as a free parameter, the whole wavelength range is 
%better reproduced. However, the  unrealistically high values of the E(B-V),
%in comparison with foreground values, are hard to explain and again suggest 
%that a passive SFH does not account for the observed SEDs.
Ages and stellar masses of the outer ring structures were estimated using FUV 
and NUV measurements because only upper limits 
can be derived from the  SDSS images. For the ring structures we 
derive very young ages ($\la$ 200 Myr) adopting a Single Stellar population 
(SSP), and $\sim$ 1 Gyr with an Elliptical SFH while assuming Spiral 
SFH ages are between 2 and 4 Gyr. 
The stellar mass of the outer rings is about 1-4\%, 1\%, 
1\% and 5-8\% of the total stellar mass of NGC 1533, NGC 2962, NGC 2974, 
NGC 4245, NGC 5636, respectively \citep[details are provided in][]{Marino10b}.

%The stellar mass of the outer rings is in the range (0.5-3.7)\%, (0.3-0.7)\%, 
%(0.7-1)\%, (5-8)\% of the total stellar mass of NGC 1533, NGC 2962, NGC 2974, 
%NGC 4245, NGC 5636, respectively.

Combining the UV and the SDSS photometry we plot in Figure~\ref{fig3} 
the  (FUV-NUV) vs. (NUV-$r$) color  diagram for ring structures (triangles), 
the total S0 ring galaxies (full dots) and ETGs without ring (open circles) 
from \citet{Marino10a}.  In the (FUV-NUV) vs. (NUV-r) plane rings  and  
entire galaxies appears well separated in color.  The two sets of
 {\tt GRASIL} models are also shown to help in interpreting the data. The color 
of the rings are  compatible either with models for a continuous 
SFH (blue lines) or with the SFH  adopted for elliptical (red lines)
but in the very young ($<$ 1 Gyr) phase.     
The global colors of the galaxies plotted in the figure 
are affected by the younger (bluer) rings
so that, on the average, the galaxy appears in all rejuvenated, in particular the cases 
of NGC 5636 and NGC 4245. Notice, however, that a large fraction
of the ETGs from the \citet{Marino10a} sample appear quite old objects.
%At the same time, the global color of the S0s with ring structure result 
%in all rejuvenated, .
 
\begin{figure}[!h]
 \includegraphics[width=\columnwidth]{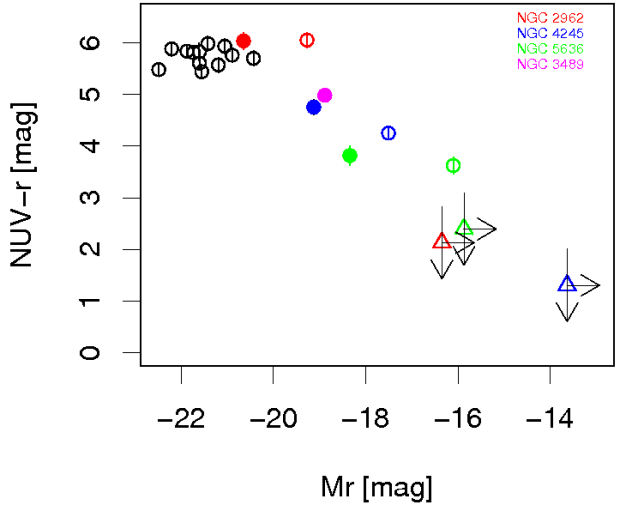} 
   \caption{$M_r$ vs $(NUV- r)$ measurements of the outer ring structures only 
(triangles), of the central portion (empty circles)  
and of the total extent  (filled circles) of NGC 2962 
(red), NGC 5636 (green),  NGC 4245 (blue) and NGC 3489 (magenta). 
Black open circles plot ETGs in  \citet{Marino10a} without  ring structures. 
NGC 2962 is  in the red sequence  \citep[e.g.][]{Salim07} 
while rings lie the blue sequence.   NGC 3489, NGC 4245 and NGC 5636 lie 
in the  so called green valley }   %% no full stop at the end of caption
\label{fig4}
\end{figure}

In Figure~\ref{fig4} we plot the (NUV-$r$) vs. M$_r$ color magnitude 
relation for our galaxies and the sample of ETGs in \citet{Marino10a}. 
ETGs define, with a large dispersion, the red sequence in this plane. 
S0s with rings like  NGC 4245 and NGC 5636 are
located  in the so called green valley, where  evolving objects are found, 
while the ring structures  are located in the  blue sequence.

% -----------------------------------------Figure 4
%----------------------------------------- end Figure 4

\section{Star Formation triggering mechanisms}

The blue, star forming rings in NGC 4245, NGC~5636 and
NGC~1533 are located at the end of the bar structure. The 
optical and UV images, shown in Figure~\ref{fig1},
suggests that NGC 2962 has two rings, one at the edge of the bar,
the other is an outer ring. Only the outer ring in NGC~2962 is
visible in UV {\it GALEX} images. 
The structure of the ring and the outer features detected
in NGC~2974 are reminiscent of the NGC 2962 case, which is seen
more face-on. We remind the presence of a shell structure in NGC~2974
revealed by \citet{Tal09}. Shell structures suggest the occurrence of an accretion 
phenomenon. 

According to different statistical estimations, the outer rings are observed in a 
significant fraction, up to 20$-$30\%, of spiral and lenticular galaxies and are
closely associated with the non-axisymmetric structures like bars, ovals or 
triaxial bulges. Distinct ring structures are often found in the same galaxy
\citep{Emsellem96}.  In terms of the theory, the empirical subdivision 
of the rings into nuclear, outer and inner corresponds to the inner, 
outer Lindblads and ultra-harmonic  dynamical resonances between 
 the epicyclic oscillations in the stellar component and rotation of the bar
\citep[see the kinematic review of][]{Moiseev09}. The literature reports cases 
of rings without bar  which are explained by weakly 
triaxial distortions, or by tidal actions from a gravitationally bounded 
companion, since the non-axisymmetric gravitational perturbations 
in these cases are similar to those from a bar \citep{Moiseev09}.

Although NGC~2974 may have suffered an accretion event, we suggest that
rings in our barred S0s sample are an effect of the secular (inner) evolution
\citep[see also][]{Jeong07}.
The tumbling bar redistributes gas angular momentum, and the gas accumulates  at the
 Lindblad's resonances \citep[see a review by][]{Buta96} where the star formation
may take place. Notice that in NGC~1533 the HI is located outside the
galaxy ring: the masses of the outer rings do not require a large gas supply,  and are 
consistent with these strcutures having formed from mass loss by the older 
population of stars \citep{vandenbosch98}.

Kinematics observations will be crucial to fully characterize the phenomenon.
 
\acknowledgments
A. Marino acknowledges support from the Italian Scientist and Scolar 
of North America Foundation (ISSNAF)
through an ISSNAF fellowship in Space Physics and Engineering, 
sponsored by Thales Alenia Space.
RR acknowledges financial support from 
the agreement ASI-INAF I/009/10/0.
{\it GALEX} is a NASA Small Explorer, launched in April
2003. {\it GALEX} is operated for NASA by California Institute of
Technology under NASA contract  NAS-98034. 
 {\tt IRAF} is distributed
by the National Optical Astronomy Observatories, which are operated
by the Association of Universities for Research in Astronomy, Inc.,
under cooperative agreement with the National Science Foundation.
This research has made
use of  SAOImage DS9, developed by Smithsonian Astrophysical 
Observatory and of the NASA/IPAC Extragalactic Database (NED) which 
is operated by the Jet Propulsion Laboratory, California Institute of
Technology, under contract with the National Aeronautics and Space
Administration.
We acknowledge the usage of the HyperLeda database (http://leda.univ-lyon1.fr).

\end{document}